 \documentclass{emulateapj}

\usepackage{natbib}

\newcounter{time}
\def\acetylene{\textrm{C}_{2}\textrm{H}_{2}}

\def\propane{\textrm{C}_{3}\textrm{H}_{8}}

\shorttitle{Titan's propane}
\shortauthors{Roe et al.}

\received{2003 August 27}
\accepted{2003 September 18}
\begin{document}


\title{Propane 
on Titan}


\author{H.G.~Roe$^{1,2}$, T.K.~Greathouse$^{1,3}$, M.J.~Richter$^{1,4}$, and J.H.~Lacy$^{1,3}$}
%\affil{California Institute of Technology, 
%Division of Geological and Planetary Sciences, Pasadena, CA 91125}
%\email{hroe@gps.caltech.edu}
%
%\author{T.K.~Greathouse$^{1,3}$}
%\affil{Department of Astronomy, 
%University of Texas at Austin, Austin, TX 78712}
%
%\author{M.J.~Richter$^{1,4}$}
%\affil{Department of Physics,
%University of California, Davis, CA 95616}
%
%\and
%
%\author{J.H.~Lacy$^{1,3}$}
%\affil{Department of Astronomy, 
%University of Texas at Austin, Austin, TX 78712}


\altaffiltext{1}{Visiting Astronomer at the Infrared Telescope Facility, which is operated by the University of Hawaii under Cooperative Agreement no.\ NCC 5-538 with the National Aeronautics and Space Administration, Office of Space Science, Planetary Astronomy Program. }
\altaffiltext{2}{O.K. Earl Postdoctoral Scholar in Planetary Science,
California Institute of Technology, 
Division of Geological and Planetary Sciences, Pasadena, CA 91125;
hroe@gps.caltech.edu}
\altaffiltext{3}{Department of Astronomy, 
University of Texas at Austin, Austin, TX 78712; 
tommyg@astro.as.utexas.edu, lacy@astro.as.utexas.edu}
\altaffiltext{4}{Department of Physics,
University of California, Davis, CA 95616; richter@physics.ucdavis.edu}

%% You can insert a short comment on the title page using the command below.


\begin{abstract}
We present the first observations of propane (C$_3$H$_8$)
on Titan that unambiguously resolve 
propane features from other numerous stratospheric emissions.
This is accomplished using a $R=\lambda/\delta\lambda\approx10^5$ 
spectrometer (TEXES)
to observe propane's $\nu_{26}$ rotation-vibration band near 748~cm$^{-1}$.
We find a best-fit fractional abundance of propane in Titan's stratosphere of
$(6.2\pm1.2)\times10^{-7}$ in the altitude range to which we
are sensitive (90-250~km or 13-0.24~mbar).
\end{abstract}




%% BE SURE TO COMMENT OUT THESE LINES BEFORE SUBMITTING
% To be submitted to ApJL;
% Draft as of $\thetime$ min after midnight $\today$. 

\section{Introduction}

Titan's thick atmosphere is simultaneously analogous to and extraordinarily 
different from that of Earth.  Both atmospheres are composed mostly of 
nitrogen.  Both atmospheres have a major component that can exist as a solid, 
liquid, or gas (water on Earth, methane on Titan).  Both have 
similar vertical temperature-pressure structures, although Titan's
atmosphere is 
$\sim$200~K cooler and is greatly extended relative to Earth due to much lower
surface gravity.  
On Titan a complicated network of photochemical reactions leads from
methane (CH$_4$) to the 
formation of numerous heavier hydrocarbons, including propane ($\propane$).
Measuring the abundances of Titan's hydrocarbons provides an important test
of models of Titan's atmospheric chemistry and, more generally, of our 
fundamental understanding of atmospheric chemical and physical processes.

Photochemical models \citep[e.g.\ ][]{1984ApJS...55..465Y, 
1995Icar..113....2T,1996JGR...10123261L,2000Icar..147..386B}
predict that propane is formed via 
%\begin{equation}\textrm{
C$_2$H$_5$ + CH$_3$ + M $\rightarrow$ C$_3$H$_8$ + M
%}\end{equation}
where CH$_3$ is created by methane photolysis and C$_2$H$_5$ by 
%\begin{equation}\textrm{
C$_2$H$_4$ + H + M $\rightarrow$ C$_2$H$_5$ + M.
%}\end{equation}
The primary removal mechanism for propane is
condensation at the cold tropopause, although 
photolytic destruction and reaction with C$_2$H consume $\sim$5-15$\%$ of the
propane formed.  
In the model of \citet{1996JGR...10123261L} propane condensed onto
Titan's surface accounts for $\sim$1$\%$ of the carbon removed from Titan's
atmosphere.

Detections of propane's $\nu_{26}$ rotation-vibration band near 748~cm$^{-1}$
 were based on Voyager~I and Infrared
Space Observatory (ISO) spectra.  Voyager spectra had 4.3~cm$^{-1}$ 
resolution ($R=\lambda/\delta\lambda=170$ at $748$~cm$^{-1}$) and 
initial claims of a propane abundance of 2$\times10^{-5}$
\citep{1981Natur.292..683M} were later reduced to 7$\pm4\times10^{-7}$
\citep{1989Icar...80...54C}.
The resolution of these spectra left every band of propane blended with
other emission features.  At 748~cm$^{-1}$ the $\nu_{26}$ band of propane
was blended with features of acetylene ($\acetylene$)
 and the derived abundance (or even 
detection) of propane is strongly dependent on the assumed $\acetylene$ 
vertical abundance profile.  Spectra taken by ISO were of higher
resolution ($R\approx1850$ at $748$~cm$^{-1}$) but were still
unable to resolve propane from acetylene.  From these data 
\citet{2003Icar..161..383C} derived a nominal propane abundance of
2$\pm1\times10^{-7}$, although they acknowledged the substantial difficulties
of disentangling propane emission from that of acetylene.

\section{Observations}

\begin{figure}
\epsscale{1.2}
\plotone{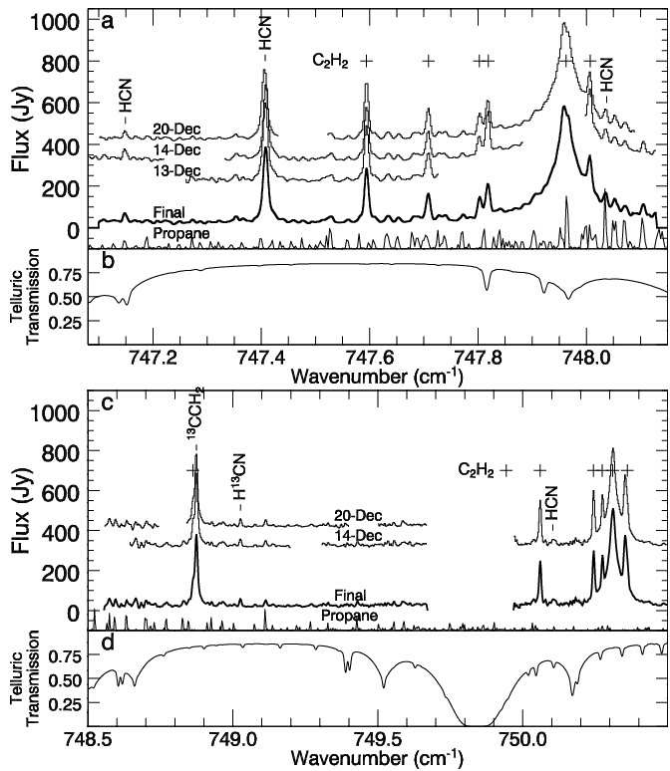}
\caption{
%\figcaption[f1.ps]{
Reduced spectra from each night (offset for clarity) 
and a final combined
spectrum are shown on the upper panels ({\bf a},{\bf c}), 
while the calculated telluric
transmission \citep[Appendix A]{roethesis} 
is shown on the lower panels ({\bf b},{\bf d}).  
When compared with an
empirical absorption coefficient spectrum of propane (shown at
bottom of {\bf a},{\bf c} in arbitrary units; See \S~3 for more details.), 
nearly all unlabeled features
in the observed spectra are seen to be associated with propane. \label{fig1}}
\end{figure}

Observations were made at the 3.0 meter NASA 
Infrared Telescope Facility (IRTF) using the TEXES mid-infrared 
high-resolution grating spectrograph \citep{2002PASP..114..153L} on 
13, 14, and 20 December 2002 UT in several overlapping spectral settings
around propane's $\nu_{26}$ fundamental vibration at 748~cm$^{-1}$.
%% \citet{herzbergVolII} if need a citation about which nu number it is
At this wavenumber a single spectral setting covers $\sim$5~cm$^{-1}$
at a resolution of $R\approx10^{5}$ with slight gaps between 
spectral orders.  The strongest $\nu_{26}$ propane lines are blocked
by telluric absorption at 748.3~cm$^{-1}$.
The diffraction limit of IRTF at these wavelengths and Titan's diameter
are both $\approx0\farcs9$.  With a slit width of $1\farcs5$ our data
represent a disk average spectrum of Titan.
The reduced spectrum from each night, corrected for the appropriate
Doppler shift, is shown in Fig.~\ref{fig1}
along with the final combined spectrum.
All data were reduced using the standard pipeline reduction
detailed in \citet{2002PASP..114..153L}.
Telluric transmission was corrected with Callisto on the 
first two nights and the Becklin-Neugebauer object \citep{1967ApJ...147..799B}
on the third night.
We flux calibrated using observations of $\beta$ Gem 
\citep{1995AJ....110..275C} on the final night, and
estimate that this calibration is accurate to 10-20\%.

\section{Radiative transfer modeling}

\defcitealias{1998JQSRT_60_665short}{Rothman et al.\ 1998}
\defcitealias{geisa1999short}{Jacquinet-Husson et al.\ 1999}

We use a new line-by-line code \citep{roethesis}, dividing Titan's atmosphere
below 1000~km into 50 spherical shells 
evenly spaced in $log\left(P\right)$ and using spectral bins of
$3\times10^{-4}$~cm$^{-1}$ in order to resolve even the narrowest emission
lines.  Line parameters for acetylene (C$_2$H$_2$) and HCN are 
from HITRAN \citepalias{1998JQSRT_60_665short}, 
with the HCN line positions adjusted to agree
with the observations of \citet{DuxburyYu1989}.  The temperature-pressure
profile is the `recommended' profile of \citet{1997hspm.conf..243Y}.
Scattering effects are ignored and Titan's haze is modeled as a single layer
with a lower cutoff of 50~km and with haze opacity scale height equal to the 
gas density scale height.  The abundances of HCN and $\acetylene$ are each
parameterized with their fractional abundances at 1~mbar 
($F_{HCN,1mbar}$, $F_{C_2H_2,1mbar}$)
and with the slope of 
$log$(abundance)-$log$(pressure) ($n_{HCN}$, $n_{C_2H_2}$), i.e.\ 
$F_{HCN} = F_{HCN,1mbar}  (P_{mbar})^{n_{HCN}}$.  
We investigated both a constant propane vertical profile as well as scaled 
versions of an abundance profile predicted by photochemical modeling 
\citep{1996JGR...10123261L,2000Icar..147..386B}.
Species are held to saturation vapor pressures below their condensation
altitudes.  In order to calculate Titan's total flux,
the model is run at 32 points from the center of Titan's disk to off the
edge of Titan's solid limb, with the calculation points more closely spaced
near the limb.  

\begin{figure}
\epsscale{1.15}
\plotone{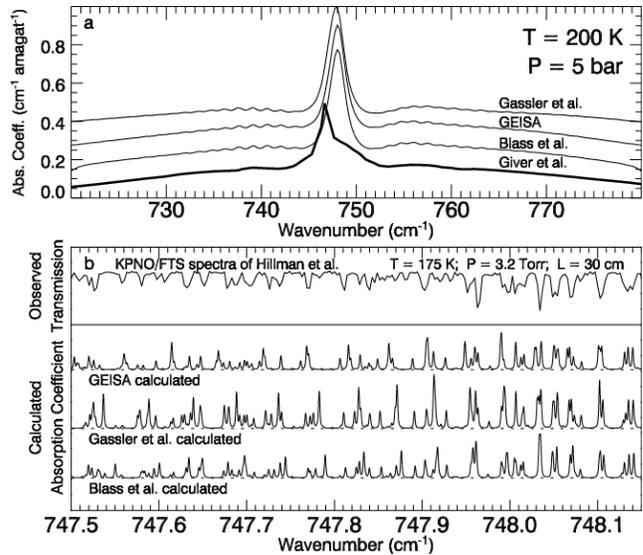}
\caption{
%\figcaption[f2.ps]{
({\bf a}) Comparison between low resolution laboratory measurements
of \citet{GiverVaranasiValero1984} at 200~K with calculated
spectra for the same conditions using the available line lists.
Calculated spectra are scaled to match the observed spectrum as closely as 
possible and offset for clarity.  There is good agreement in the wings
of the band, however over $\sim$740-760~cm$^{-1}$ there is very poor
agreement between calculations and observation.
({\bf b}) Comparison of calculated spectra using the available line lists
at 175~K and 3.2 Torr
with an observed transmission spectrum of propane at the same conditions
from \citet{HillmanReuterJennings1992}.    
There is good agreement at some wavenumbers (e.g.~748.03-748.15~cm$^{-1}$) and
poor agreement at others (e.g.~$<748.0$~cm$^{-1}$).
 \label{fig2}}
\end{figure}

The modeling of propane's emission spectrum requires additional
discussion.  Several linelists exist for propane's $\nu_{26}$ band at
748~cm$^{-1}$.  These include: the GEISA databank
\citepalias{geisa1999short}; an unpublished list based on fits to the
laboratory spectra of \citet{NadlerJennings1989} and
\citet{HillmanReuterJennings1992} (Blass et al.\ 1988); and a list
generated using the code of \citet{Typke1976} and the parameters of
\citet{GasslerReissenauerHuttner1989}.  Figure~\ref{fig2} shows
calculated spectra using each of these linelists compared to
low-resolution laboratory spectra of \citet{GiverVaranasiValero1984}
and to the high-resolution spectra used in
\citet{HillmanReuterJennings1992}.  Given the poor fit of the
calculated spectra to the observed spectra in Fig.~\ref{fig2} at the
wavenumbers we observed, we opted to use the low-pressure (1-3~Torr)
low-temperature ($\sim$175~K) high-resolution (R$\approx3\times10^5$)
transmission spectra of
\citet{HillmanReuterJennings1992}\footnote{Available at
http://diglib.nso.edu/nso\_user.html} to construct an empirical
absorption coefficient ($k_{\nu}$) spectrum for propane.  This
involved fitting for the smoothly varying baseline and then converting
to $k_{\nu}$ units (cm$^{-1}$ amagat$^{-1}$) using the sample's known
pathlength (30~cm), pressure, and temperature.  In using this
empirical $k_{\nu}$ spectrum we are implicitly assuming that $k_{\nu}$
is independent of temperature.  Our propane observations primarily
probe regions of Titan's atmosphere that are at 135 to 175~K and are
therefore reasonably well matched by the conditions of these
laboratory data used to construct $k_{\nu}$.  Most of our spectral
fitting was performed using this empirical $k_{\nu}$ for propane.  As
a test of the importance of the mismatch between Titan atmospheric and
laboratory temperatures we re-fit our data using the Blass et al.\
linelist scaled to fit our empirical $k_{\nu}$ at 175~K, along with
the partition sum parameterization of \citet{FischerGamache2002a}, to
represent propane.

\section{Fitting model to data}
 
\begin{figure}
\epsscale{1.2}
\plotone{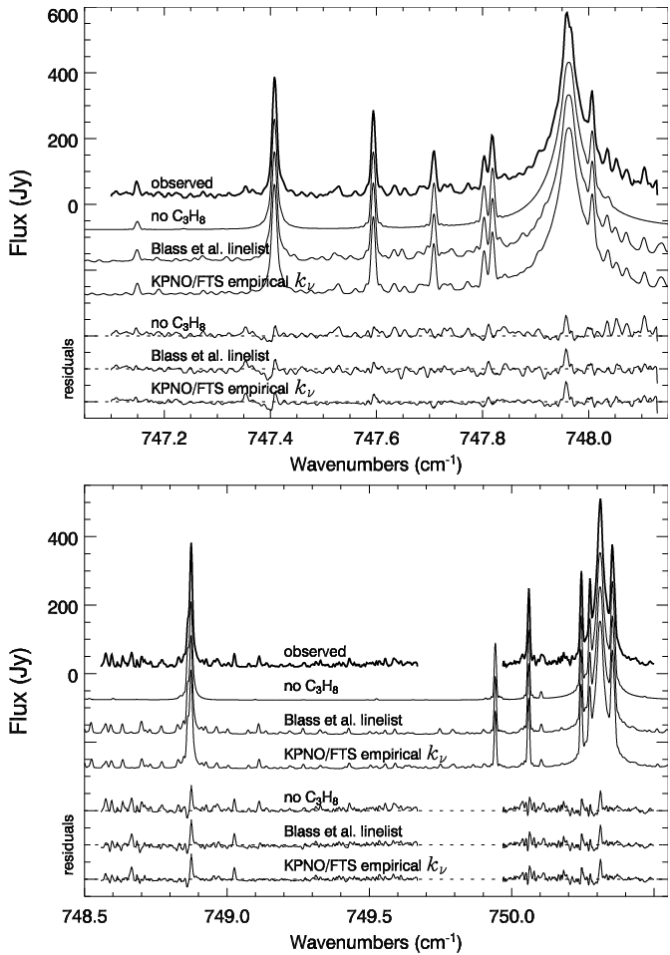}
\caption{
%\figcaption[f3.ps]{
Comparison of observed spectrum with model convolved to the observed 
spectral resolving power.
Spectra are offset for clarity and residuals 
are shown on the same scale.
The best fit is obtained using our empirical $k_{\nu}$ spectrum for propane.
 \label{fig3}}
\end{figure}

We fit model to data with an `amoeba'-type minimization
while varying six parameters: total haze opacity ($\tau_{haze}$),
propane ($F_{C_3H_8}$), 
acetylene ($F_{C_2H_2,1mbar}$, $n_{C_2H_2}$), and HCN
($F_{HCN,1mbar}$, $n_{HCN}$).  
Figure~\ref{fig3} compares the observed
spectrum with the best-fit model with and without propane, in one
case using the Blass et al.\ linelist and in the other using our
empirical $k_{\nu}$ to represent propane.  The
detection of propane is unambiguous and the best-fit abundance assuming
a constant vertical profile and using our empirical $k_{\nu}$ is 
$6.2\times10^{-7}$.  For acetylene and HCN our best fits are characterized by 
$F_{C_2H_2} = 4.4\times10^{-6} (P_{mbar})^{-0.45}$ and 
$F_{HCN} = 3.8\times10^{-7} (P_{mbar})^{-0.18}$.  The haze is best
fit by $\tau_{haze} = 0.21$, although this value is extremely dependent on
the assumed vertical haze profile and should not be over interpreted.

We found that reasonable fits to the
HCN and $\acetylene$ features required a non-constant vertical abundance.
Even with our two variable log-log parameterization we see non-zero
residuals on these features, suggesting a more complicated
vertical structure which we will explore further in future work with
similar spectra covering many more transitions. Other significant
 non-zero residuals are explained as: an unidentified species 
 at 747.35~cm$^{-1}$,
an imperfectly corrected telluric feature at 748.68~cm$^{-1}$, 
not fitting for the $^{13}$CCH$_2$/C$_2$H$_2$ ratio at 748.88~cm$^{-1}$, 
and
an H$^{13}$CN feature at 749.02~cm$^{-1}$ that is missing from the model.

Our focus here is propane and thus we do not present uncertainties on the
vertical abundance parameters for HCN and $\acetylene$.  As is
apparent in Fig.~\ref{fig1}, random errors are negligible in our spectra.  The
other sources of uncertainty are: flux calibration, errors associated
with our empirical absorption coefficient method, 
the extent to which propane abundance is
not constant with altitude, the extent to
which Titan's temperature-pressure profile deviates from that derived
by \citet{1997hspm.conf..243Y}, and the extent to which
propane's abundance varies with latitude.

The uncertainty in flux calibration is at worst 20\%, which maps
almost linearly into a 20\% uncertainty in propane abundance.  The primary
potential error from our empirical absorption coefficient method is the
mismatch of laboratory sample temperature (175~K) to the temperature of the
propane in Titan's stratosphere.
In our model spectra $>80$\% of the propane flux arises from 
 the pressure range 13-0.24~mbar (90-250~km altitude), 
where temperatures range over 135-175~K.  To 
investigate this uncertainty we re-fit using the Blass et al.\ linelist
(See Fig.~\ref{fig3}.).  The overall fit is lower quality than with
our empirical $k_{\nu}$ and the best-fit propane abundance is only
1.1$\pm$0.1 times more.

\begin{figure}
\epsscale{1.1}
\plotone{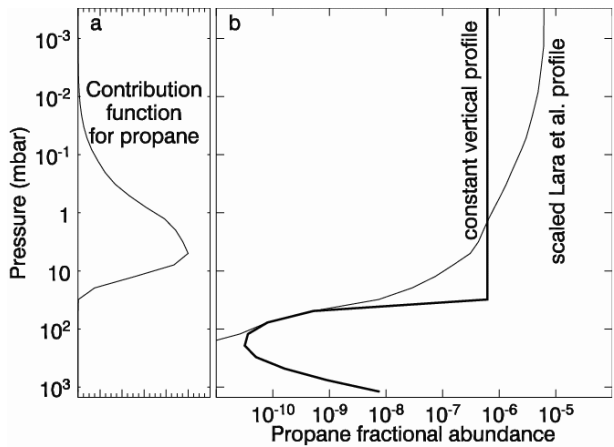}
\caption{
%\figcaption[f4.ps]{
({\bf a}) Contribution function of propane emission at
the middle of the 748.10~cm$^{-1}$ feature.  The 
contribution function is 
the partial derivative of propane emission with respect to $\log P$.
({\bf b}) Comparison of the two best-fit vertical abundance profiles.
 \label{fig4}}
\end{figure}

Recent photochemical models 
\citep[e.g.][]{1996JGR...10123261L,2000Icar..147..386B} predict that the 
propane abundance is relatively constant over 400-850~km (11-1.5~$\mu$bar).  
Above these
altitudes propane is depleted by photolysis and reaction with C$_2$H,
while below this range the effect of eddy diffusion and the cold-trap of the
tropopause is to reduce the propane abundance.
Our observations are primarily 
sensitive to the pressure range 13-0.24~mbar (90-250~km altitude), 
as shown in Fig.~\ref{fig4}a.
We re-fit our spectra using the
model predicted propane vertical profile and found a best fit when the
model predicted profile is multipled by a factor of 2.9.
Figure~\ref{fig4}b shows this best-fit scaled profile of 
\citet{1996JGR...10123261L} along with the best-fit constant
vertical abundance profile.  The quality of the fits to our data
do not distinguish between these two cases.

The uncertainty in the \citet{1997hspm.conf..243Y} temperature-pressure
profile is not well quantified, and therefore our reported uncertainties
do not account for this source of error.  Roughly, a change in
stratospheric temperature of $\pm$5~K requires a $\mp$20$\%$ change in
propane abundance.

There is no strong evidence that propane abundance is constant with
latitude on Titan.  A first attempt at modeling the seasonal variation
of photochemistry on Titan \citep{2001Icar..152..384L} predicts that
propane abundance will vary by a factor of a few from north to south
depending on season.  With TEXES on an 8-10 meter telescope we would be able
to measure the abundance as a function of latitude, which will place
new strong constraints on models of seasonal photochemistry.

\section{Summary}

We present the first spectrally resolved 
detection of propane in Titan's atmosphere.  We measure propane's fractional 
abundance to be $(6.2\pm1.2)\times10^{-7}$, assuming the recommended
temperature-pressure profile of \citet{1997hspm.conf..243Y} and a
constant abundance with altitude and latitude.  Our observations are primarily 
sensitive to
13-0.24~mbar (90-250~km altitude).  Alternatively, our data require
the predicted propane profile of \citet{1996JGR...10123261L} to be increased
by a factor of 2.9.

The current theoretical understanding of propane's vibration-rotation spectrum
is inadequate and available linelists do not fit our Titan spectra
nor laboratory propane spectra well.  Therefore we fit for Titan's
propane abundance using an empirical absorption coefficient spectrum
derived from low temperature laboratory spectra of propane.

The most important observational advance that can be made 
towards detecting new species in planetary atmospheres is to increase the
spectral resolving power to $R>$few$\times10^4$ such that individual
lines are resolved and separated.  
High spectral resolution allows us to separate
the contribution of a minor species (propane) from a strongly emitting
species (acetylene).  


%% Included in this acknowledgments section are examples of the
%% AASTeX hypertext markup commands. Use \url without the optional [HREF]
%% argument when you want to print the url directly in the text. Otherwise,
%% use either \url or \anchor, with the HREF as the first argument and the
%% text to be printed in the second.

\acknowledgments

We thank several people who shared their time and advice, including
especially L.~Lara, B.~Blass, G.~Bjoraker, R.~Gamache, V.~Typke, and
the referee.
Observations with TEXES are 
supported by NSF grant AST-0205518.  TKG and MJR were supported by the
SOFIA project through USRA grant 8500-98-005.

\end{document}